\documentclass[twocolumn]{emulateapj}
\usepackage{natbib}

\usepackage{xspace}

\newcommand{\fig}{Figure }
\newcommand{\msun}{\,M$_{\odot}$\xspace}

\newcommand{\kms}{\,kms$^{-1}$\xspace}
\newcommand{\tab}{Table }

\newcommand{\E}{\times 10}

\newcommand{\hco}{HCO$^+$\xspace}
\newcommand{\n}{N$_2$H$^+$\xspace}

\begin{document}
\shorttitle{Signatures of Dynamical Collapse during High Mass Star Formation}
\shortauthors{Smith, Shetty, Beuther, Klessen \& Bonnell}

\title{Line Profiles of Cores within Clusters: II Signatures of Dynamical Collapse during High Mass Star Formation}
\author{Rowan J. Smith$^{1}$, Rahul Shetty$^{1}$, Henrik Beuther$^{2}$, Ralf S. Klessen$^{1}$, Ian A. Bonnell$^{3}$}
\affil{1. Zentrum f\"ur Astronomie der Universit\"at Heidelberg, Institut f\"ur Theoretische Astrophysik, Albert-Ueberle-Str. 2, 69120 Heidelberg, Germany \\
2. Max-Planck-Institute for Astronomy, K\"onigstuhl 17, 69117 Heidelberg, Germany \\
3. SUPA, School of Physics and Astronomy, University of St Andrews, North Haugh, St Andrews, Fife KY16 9SS}

\email{rowan@uni-heidelberg.de}
 

\begin{abstract}
Observations of atomic or molecular lines can provide important information about the physical state of star forming regions.  In order to investigate the line profiles from dynamical collapsing massive star forming regions (MSFRs), we model the emission from hydrodynamic simulations of a collapsing cloud in the absence of outflows.  By performing radiative transfer calculations, we compute the optically thick \hco and optically thin \n line profiles from two collapsing regions at different epochs.  Due to large-scale collapse, the MSFRs have large velocity gradients, reaching up to 20 km s$^{-1}$ pc$^{-1}$ across the central core.  The optically thin lines typically contain multiple velocity components resulting from the superposition of numerous density peaks along the line-of-sight. The optically thick lines are only marginally shifted to the blue side of the optically thin line profiles, and frequently do not have a central depression in their profiles due to self-absorption.  As the regions evolve the lines become brighter and the optically thick lines become broader.  The lower order \hco (1-0) transitions are better indicators of collapse than the higher order (4-3) transitions. We also investigate how the beam sizes affect profile shapes. Smaller beams lead to brighter and narrower lines that are more skewed to the blue in \hco relative to the true core velocity, but show multiple components in \n. High resolution observations (e.g. with ALMA) can test these predictions and provide insights into the nature of MSFRs.
\end{abstract}


\section{Introduction}
The birth of massive stars is inextricably linked to that of cluster formation, as most massive stars form at the centre of dense molecular clouds. The structure of such regions is dense and filamentary, far removed from simple spherical models. In \citet{Smith12a}, hereafter Paper I, we investigated the line profiles that would be observed from collapsing cores embedded within filaments from simulations of clustered star formation. We found that the dense filaments frequently obscured the collapsing core and a blue asymmetric collapse profile \citep{Zhou92,Walker94,Myers96} was observed in less than 50\% of sightlines. In the current paper we extend this analysis to study the line profiles produced during the initial collapse of massive star forming regions (MSFRs).

There have been numerous observational studies of line profiles from MSFRs \citep[e.g.][]{Wu03,Fuller05,Wu07,Sun09,Chen10,Csengeri11}. Generally such studies have focussed on finding infall candidates by identifying blue asymmetries in their optically thick lines ( see Paper I or the review by \citealt{Evans99}.) A common way of classifying such surveys is using the blue excess seen across the survey i.e. the number of blue biased profiles minus the number of red biased profiles divided by the total number of observations \citep{Mardones97}. Typical values for such surveys are around 10-30\%. This finding could be interpreted in several ways. First, the majority of MSFRs could be quasi-equilibrium objects containing static massive pre-stellar cores that have not yet collapsed \citep{Tan06}. Alternatively, most massive star forming regions could be collapsing according to their dynamical timescale \citep{Elmegreen00,Vazquez-Semadeni05} but the observational signature produced differs from that predicted by simple spherical models. There are alternative models of massive star formation in which massive stars are formed from well defined massive pre-stellar cores supported by super-sonic turbulence \citep{McKee03}
, however predictions of the observed line profiles resulting from such theories have not yet been calculated.

In this paper we use the simulations presented in \citet{Smith09b}, hereafter S09, in which a cluster of low mass stars are formed around massive ones to investigate the observational signatures of a collapsing MSFR. We consider the case where the MSFR has already formed a protostar at its centre which is growing rapidly in mass through accretion. This simulation follows the competitive accretion formalism \citep{Bonnell03,Bonnell04,Bonnell06} in which massive stars are formed at the centre of a collapsing cluster where the cluster potential funnels gas towards the massive protostar (see also \citealt{Klessen00,Girichidis11}). 

These simulations lack outflows, which are known to affect \hco line profiles \citep{Rawlings04}. To this end we will focus our modelling on the earliest stages of the collapse before outflows become significant. Without first considering the simple dynamical case without inflows, it would be impossible to disentangle the two effects. Despite this, we will compare our results to observational studies, that may have outflows present, in order to understand to what extent the dynamics of a complex clustered star formation region can explain observed line profiles.

Our paper is structured as follows. In Section \ref{method} we outline our method and then in Section \ref{results} we present our results. Section \ref{results} is divided into subsections, each of which identifies a key feature of the simulated observations and then directly compares the profiles to observations. We discuss some qualifications and compare to low mass star formation in Sections \ref{qual} and \ref{comparison}. Section \ref{conclusions} provides a summary.

\section{Method}\label{method}
As in Paper 1 we use the radiative transfer code RADMC-3D \citep{Dullemond12} to calculate the line profiles from star forming regions extracted from the simulations of S09. These simulations followed the evolution of a $10^4$ \msun cylindrical giant molecular cloud as it undergoes star formation using the smooth particle hydrodynamic SPH method \citep{Monaghan92}. Several massive stars were formed at the centre of clusters. Our line transfer utilises the large velocity gradient proposed by \citet{Sobolev57}, with the inclusion of ``doppler catching'' to interpolate under-resolved velocities as implemented in RADMC-3D by \citet{Shetty11,Shetty11b}. In this section we highlight only the differences in our method from the previous paper, for a full description of our methods we refer to Paper 1.

The modelled regions are chosen from \citet{Smith09b} by finding the most massive sink particles and then selecting a 0.4 pc radius region around them. Sink particles represent sites of star formation and are formed from high density gravitationally bound gas in the simulation \citep{Bate95}. The sink particles have a radius of 20 AU and it is possible that within this a multiple system is formed, however we shall use the term sink and protostar synonymously in what follows. We identify two independent regions, one of which forms a very massive sink particle ($\sim 30$ \msun) and the other a less massive sink ($\sim 10$ \msun).  

One of our chosen species, \hco, has been shown to be present in outflows \citep{Rawlings04,Paron12}. As we lack this physics in our simulation we concentrate on the early evolution of the MSFRs before outflows become dominant. We discuss potential implications of outflows in Section \ref{outflows}. 

The regions are considered at two epochs during their evolution, when the central sources are around 0.5 \msun and 5.0 \msun, corresponding to early and more evolved phases. The most massive sink reaches a final mass of around 30 \msun when the simulation is terminated. \tab \ref{ICs} outlines the properties of the two star forming regions. In addition to the central massive protostar multiple additional sites of star formation are also contained within the regions.

\begin{table}
	\caption{The mass contained within each region when the central sink mass is of order 0.5 \msun and 5.0\msun.}
	\centering
		\begin{tabular}{c c c c c }
		\hline
		\hline
		Region & Central Sink & Gas Mass & Total number & Final Sink  \\
		& Mass [\msun] & [\msun] & of sinks & Mass [\msun] \\
	 	\hline
		A & 0.5 & 370 & 36 & 29.2 \\
		B & 0.7 & 257 & 11 & 10.7 \\
		\hline
		A & 5.4 & 354 & 73 & 29.2 \\
		B & 5.0 & 260 & 18 & 10.7 \\
		\hline
		\hline
		\end{tabular}
	\tablecomments{The gas mass excludes the mass in sinks. The final sink mass corresponds to the mass of the central sink when the simulation is terminated.}
	\label{ICs}
\end{table}

In contrast to Paper 1 where we considered HCN as the optically thick tracer, for the massive star forming regions (MSFRs) we use \hco due to the greater prevalence of observational studies of massive star formation using this species \citep[e.g][]{Fuller05,Csengeri11}. We adopt a constant abundance of $A_{\mathrm{HCO}^+}=5.0\E^{-9}$ relative to the H$_2$ number density \citep{Aikawa05}. For the optically thin tracer we adopt an \n abundance $A_{\mathrm{N2H}^+}=10^{-10}$ relative to H$_2$ \citep{Aikawa05}, as in Paper I. We focus our analysis on the isolated hyperfine component (F101-012) at 93176.2527 MHz, which is displaced by 2.297 MHz from the neighbouring hyperfine lines \citep{Keto10}. Initially, we consider only the (1-0) transitions but later in the paper we shall compare to higher-order optically thick lines that trace higher densities due to their larger critical density. These two species are easily observable with the Atacama Large Millimeter Array (ALMA).

\begin{figure}
\begin{center}
\includegraphics[width=3in]{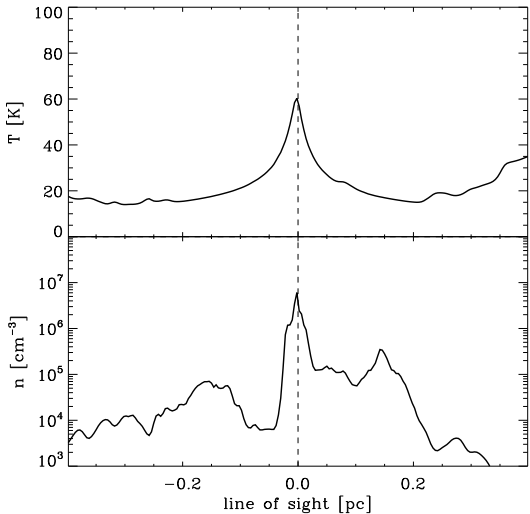}
\caption{The gas temperature along a line of sight directly through the central source when it has a mass of 5.0 \msun integrated over a 0.06pc fwhm beam. The temperature rises at the centre due to heating from the sink particle representing the massive protostar.}
\label{temp}
\end{center}
\end{figure}

There is some suggestion that in real molecular clouds \hco and \n abundances may be anti-correlated as \hco is depleted onto dust grains at low temperatures \citep[e.g.][]{Jorgensen04}. However, in massive star formation regions the temperatures around the central protostar are of order 20K or higher, so freeze out should not be a major factor. In the simulated cloud, gas temperatures are calculated using a barytropic equation of state with an additional heating term from sinks based on the YSO models of \citet{Robitaille06}. The method is described in full in \citet{Smith09b}. \fig \ref{temp} illustrates the effect of this heating along the simulated beam in the extreme case when the central sink in Region A has a mass of 5.0 \msun. Along the line of sight the temperature is highest at the peak density where the sink particle heats the gas, but also increases towards the outside of the region where the gas is diffuse and would be heated by external radiation. As emission is dependent on both density and temperature, the line profiles are dominated by emission from the central source.

All the line profiles are integrated over a Gaussian beam. For our fiducial case we consider a beam with a full-width-half-maximum (fwhm) of 0.06 pc to allow for comparison with the observational study of a high-mass starless gas clump by \citet{Beuther13}, however we shall consider the effect of altering the beamsize later in the paper.

\section{Results}\label{results}
\subsection{Velocity Fields}\label{velocities}

\subsubsection{Simulation}
For a full analysis of the dynamics of the MSFRs see S09. Here, we provide an overview of the underlying velocity fields in the simulated MSFRs.

\begin{figure*}
\begin{center}
\begin{tabular}{c c}
\includegraphics[width=3.5in]{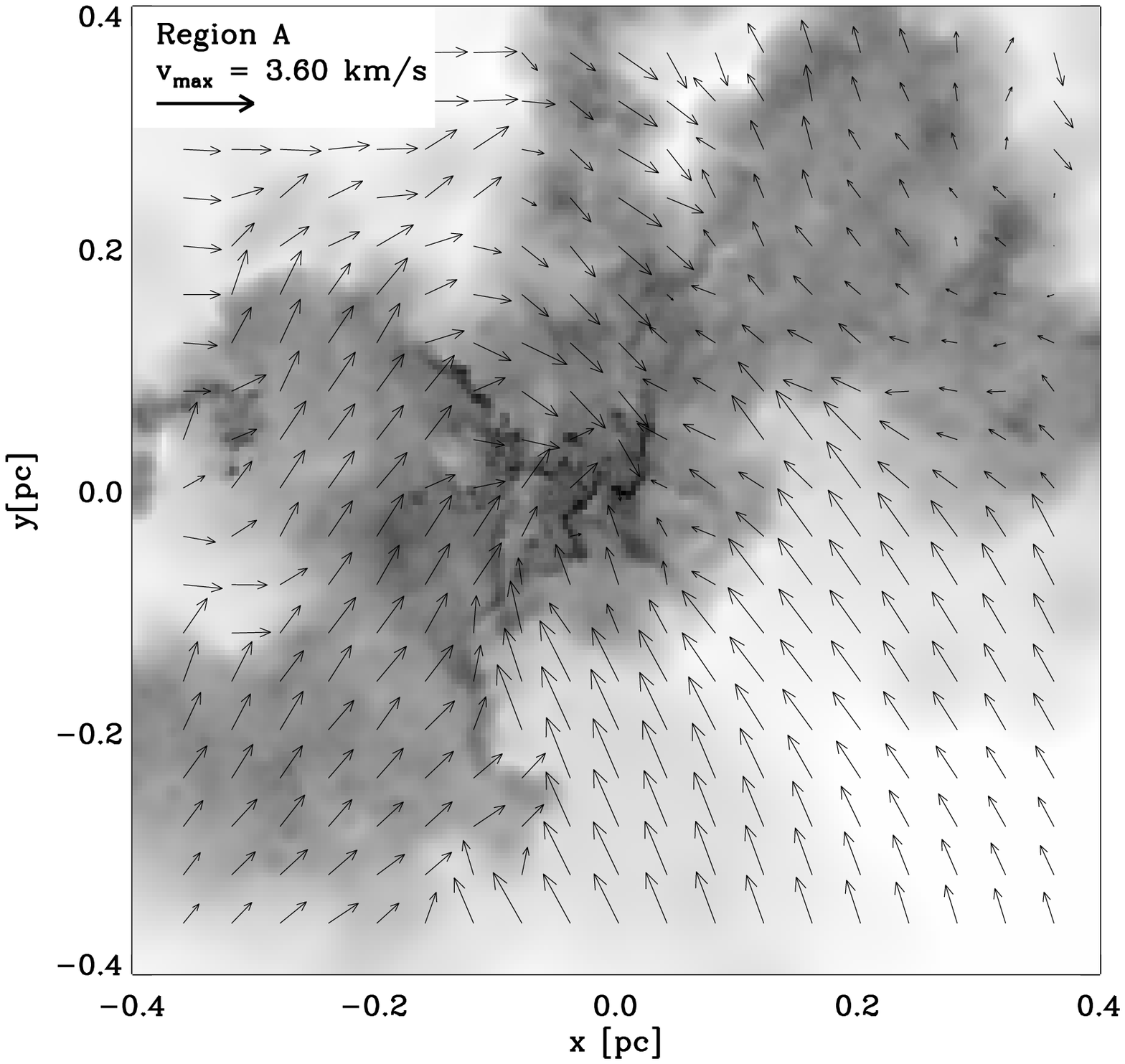}
\includegraphics[width=3.5in]{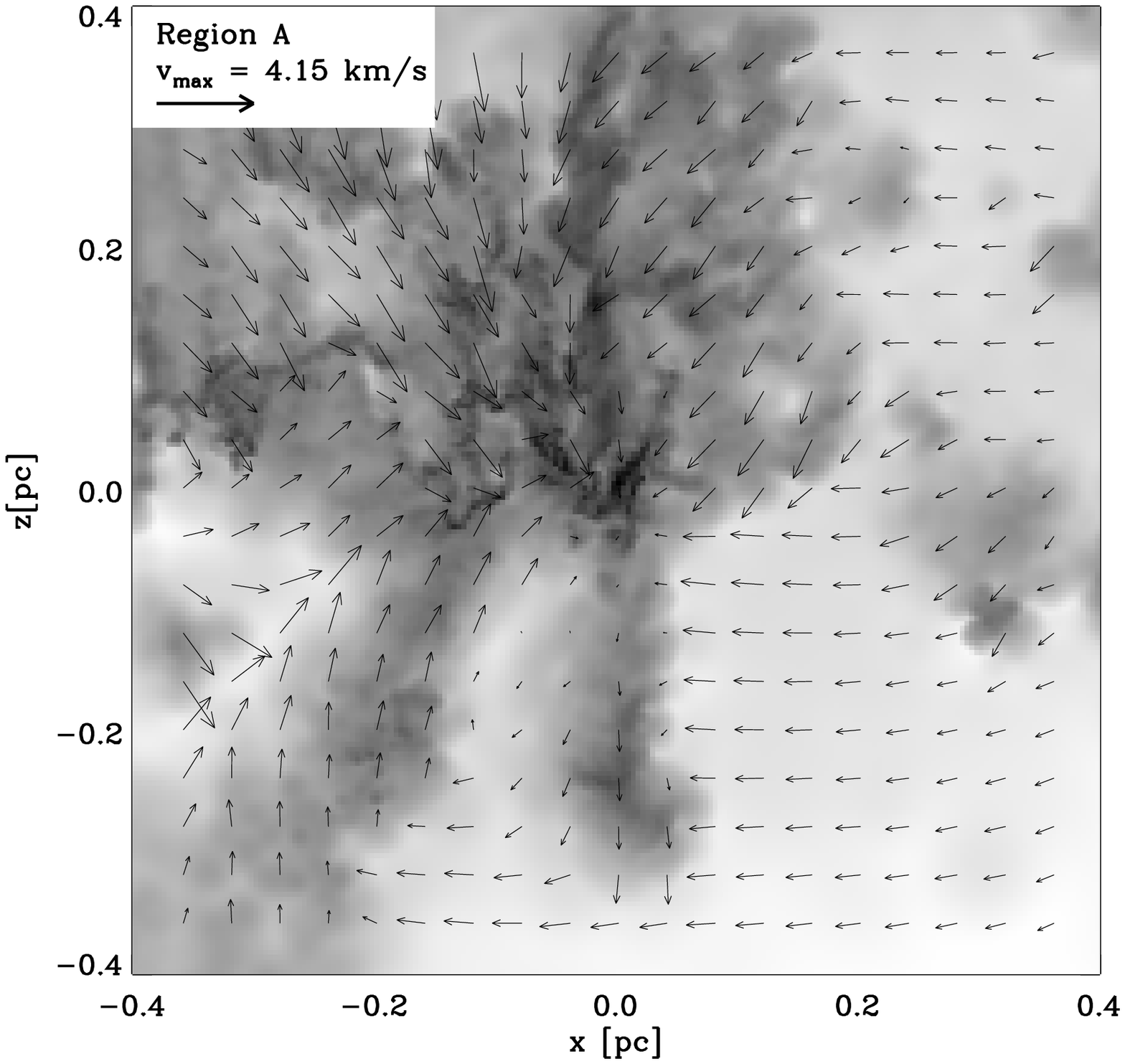}\\
\includegraphics[width=3.5in]{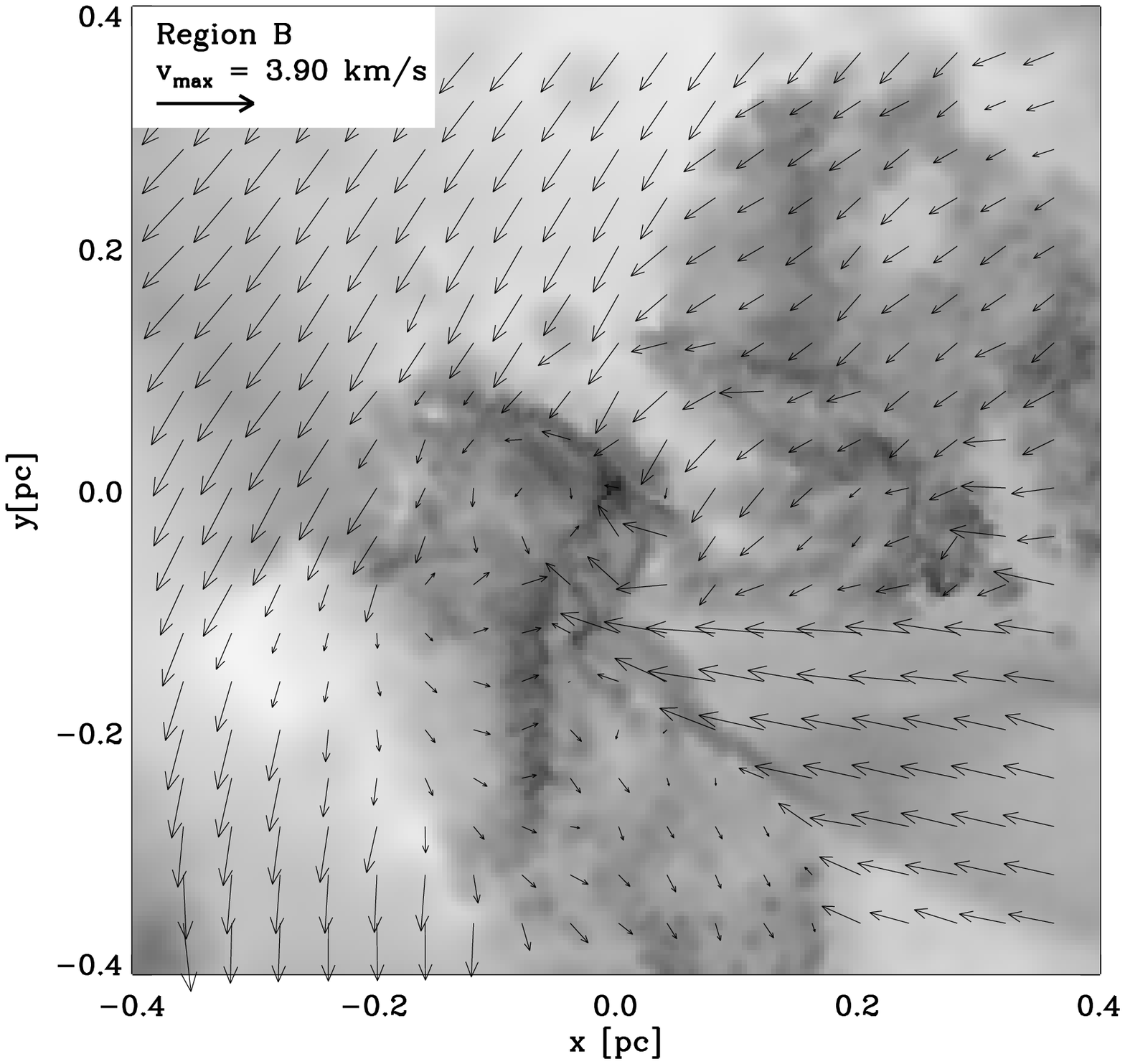}
\includegraphics[width=3.5in]{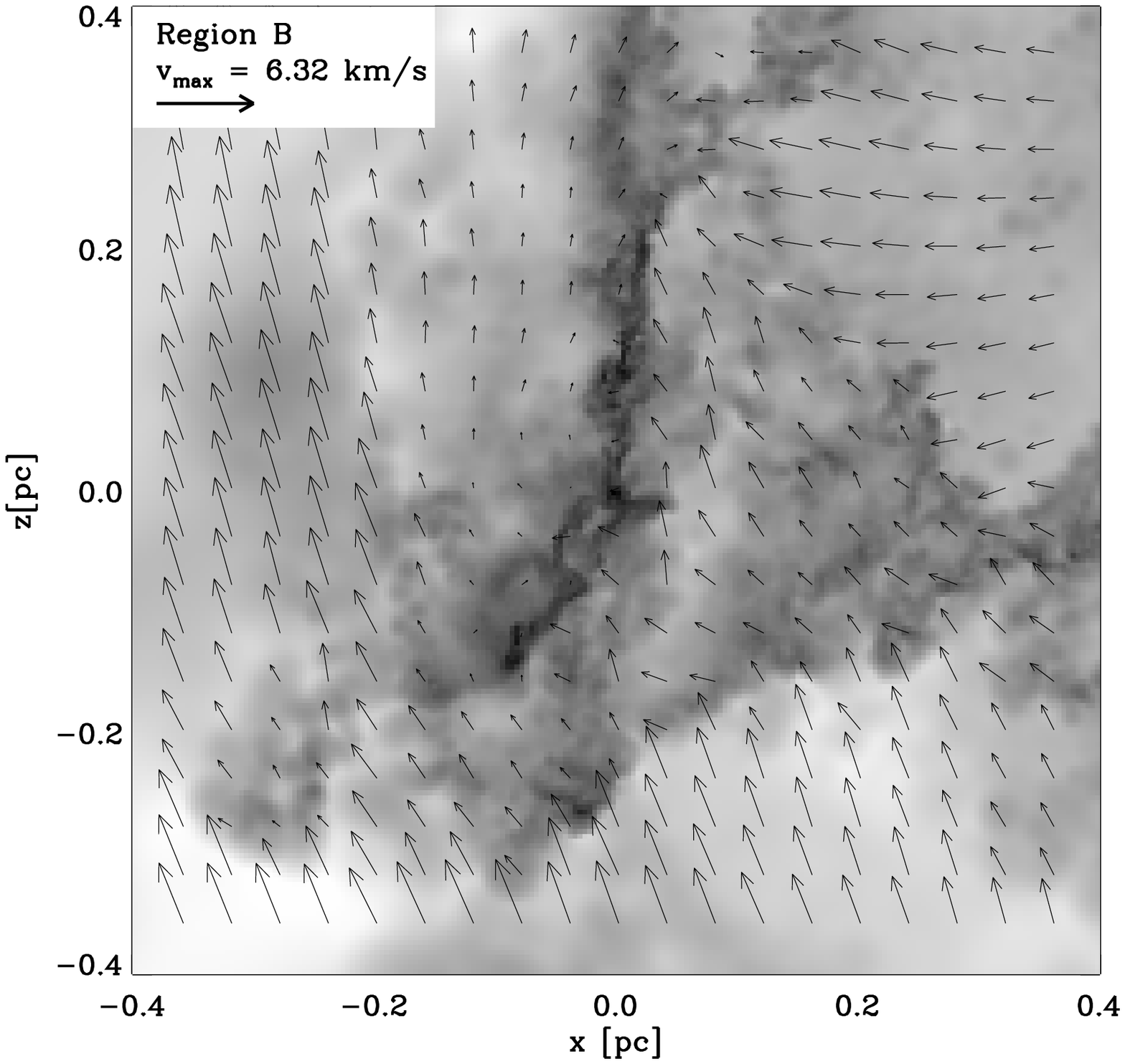}\\
\end{tabular}
\caption{The density \textit{(grayscale)} and velocity field \textit{(vectors)} of two slices through the central source in Regions A and B. In region A there is a large-scale velocity gradient towards the central massive star. In region B this is also true of the majority of the dense gas but there are also indications of the background velocity flows that formed the dense filaments in which the massive star is embedded. }
\label{velfield}
\end{center}
\end{figure*}

\fig \ref{velfield} shows the density and velocity in two slices through Regions A and B. The fields are normalised so that the most massive protostar has a position and velocity of zero. The main feature of the velocity fields is a large scale infall motion toward the central object. As discussed in S09 it is this large scale collapse of gas towards the central object that allows a massive star to form. However there are a few additional features to note. In Region A, the massive star forms at the centre of a network of converging filaments, as predicted by \citet{Myers11}, and the velocity vectors follow the contours of the filament towards the centre. In Region B there is evidence of the original velocity field of converging flows which formed the filaments in which the MSFR is embedded. The velocity field in the MSFR is therefore a combination of large scale turbulent motions and gravitational collapse. This leads to a coherent velocity field over the box, which is in strong contrast to the lower mass star formation studied in Paper 1 where the velocity field is extremely inhomogeneous. Additionally, it should be noted that unlike conventional models of spherically collapsing cores there is no envelope of static gas that surrounds the region.

\begin{figure}
\begin{center}
\includegraphics[width=3in]{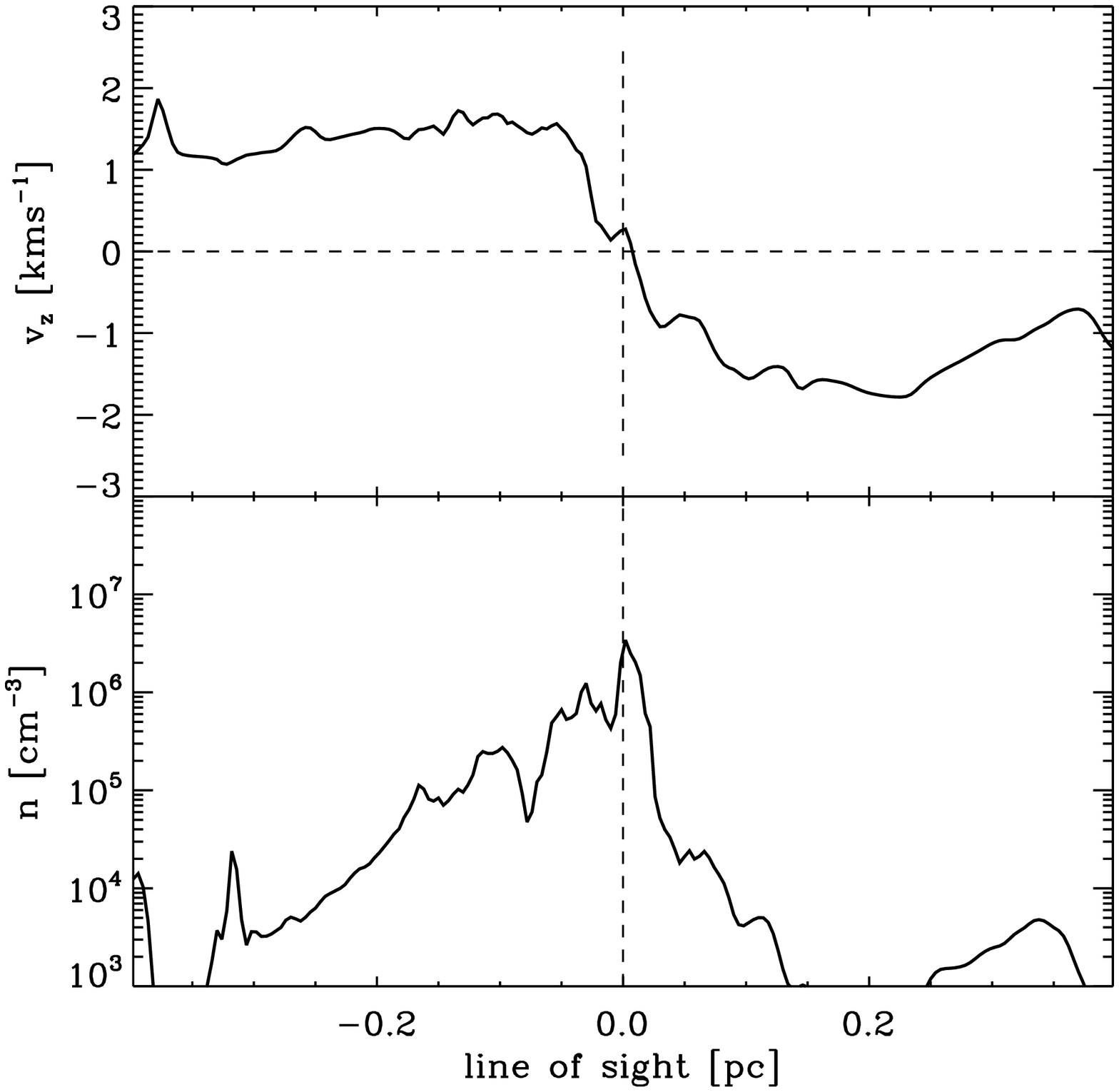}
\caption{The gas density and velocity along a line of sight directly along the x-axis of Region A. The quantities are averaged over a gaussian beam of 0.06pc fwhm. There is a large velocity gradient over the point where the massive star forms, and the density field shows considerable substructure.}
\label{beam_density}
\end{center}
\end{figure}

However, while the velocity field may be relatively coherent, the density field is much less so. As previously noted, the MSFR is formed at the convergence point of dense filaments and contains multiple sites of star formation, each of which is associated with gas over-density. \fig \ref{beam_density} shows the density and velocity field along the x-axis of Region A. There are multiple peaks in the density field each of which contributes to the overall emission separately. Local maxima in the velocity field are associated with sites of small-scale collapse, however these are superimposed on top of a larger supersonic flow of a few \kms towards the core centre from each direction. This velocity flow abruptly changes from positive to negative over the centre of the region, where the massive star forms, with a gradient of more than 2\kms occurring in less than 0.1 pc.

\subsubsection{Observations}
Evidence for large scale infall in massive star formation has been presented by a number of authors. In particular, \citet{Motte07} carried out a survey of Cygnus X searching for massive pre-stellar cores. The authors did not find any truly pre-stellar cores but instead observed large scale supersonic flows directed towards the centres of suspected young MSFRs. Further, \citet{Peretto06,Peretto07} found that the massive cluster forming clump NGC 264-C was collapsing along its axis in accordance with its dynamical timescale, and therefore channelling mass towards the Class 0 object at its centre. The violent collapse of gas to form high mass protostellar objects was also proposed by \citet{Beuther02} to explain their observed line-widths and multiple velocity components.

Another similarity between these simulations and observations is the prevalence of filamentary structure. A recent study by \citet{Peretto12} of the Pipe nebula found the only indication of star cluster formation occurred at a point of convergence of multiple filaments. Similarly \citet{Schneider12} found cluster formation at the junction of filaments in the Rosette molecular cloud.


\begin{figure*}
\begin{center}
\begin{tabular}{c c}
\includegraphics[width=3.6in]{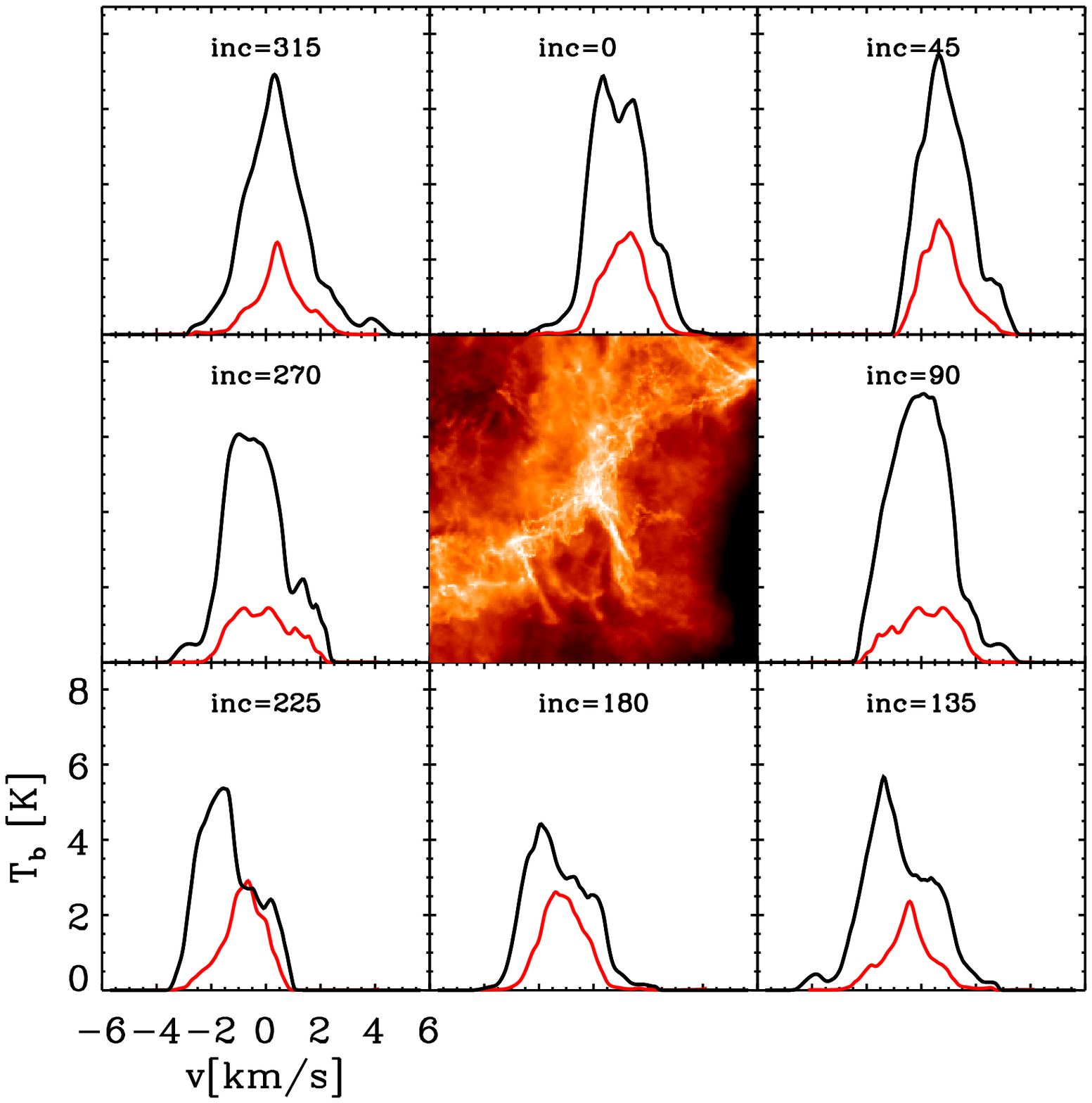}
\includegraphics[width=3.6in]{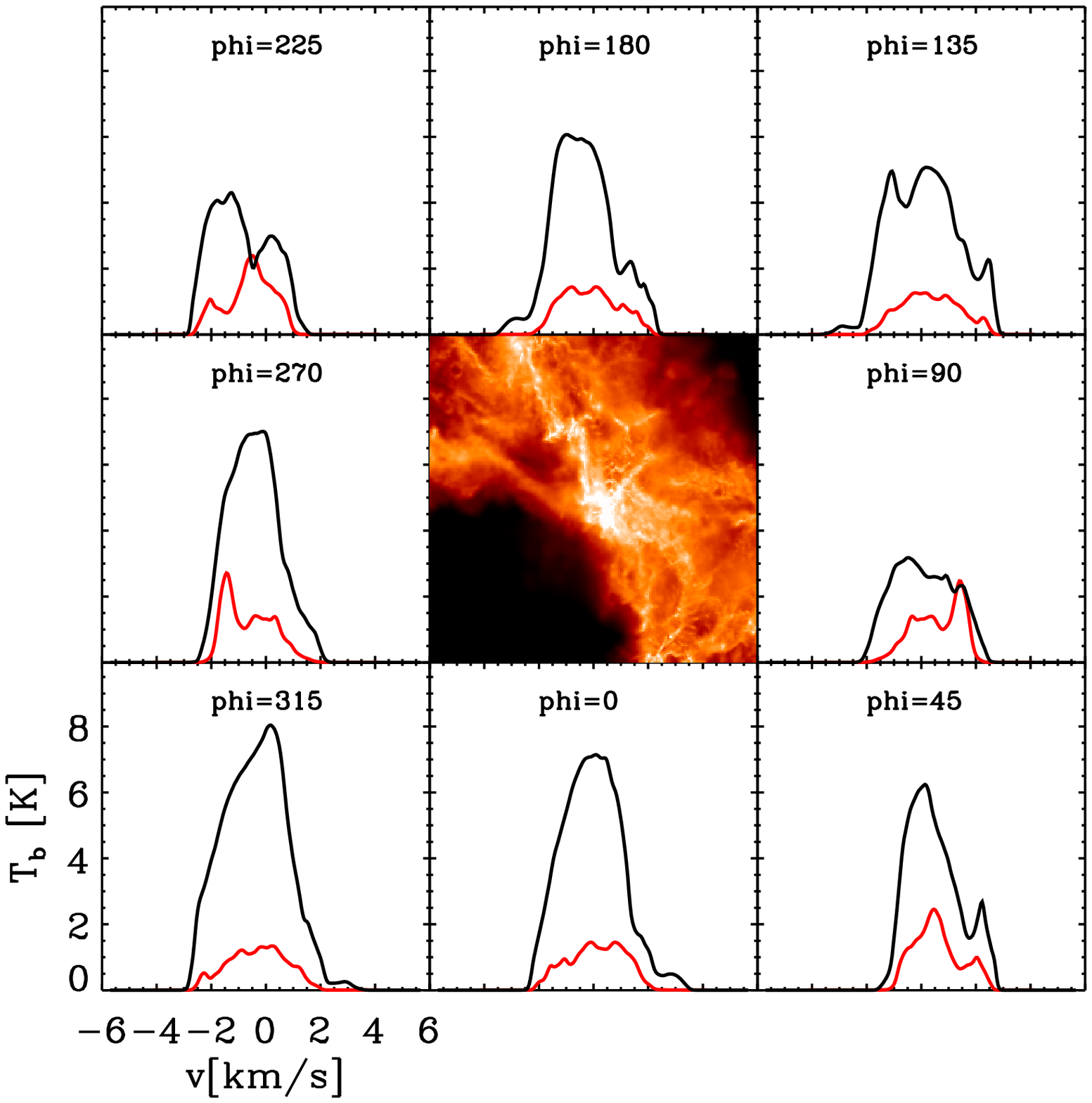}\\
\includegraphics[width=3.6in]{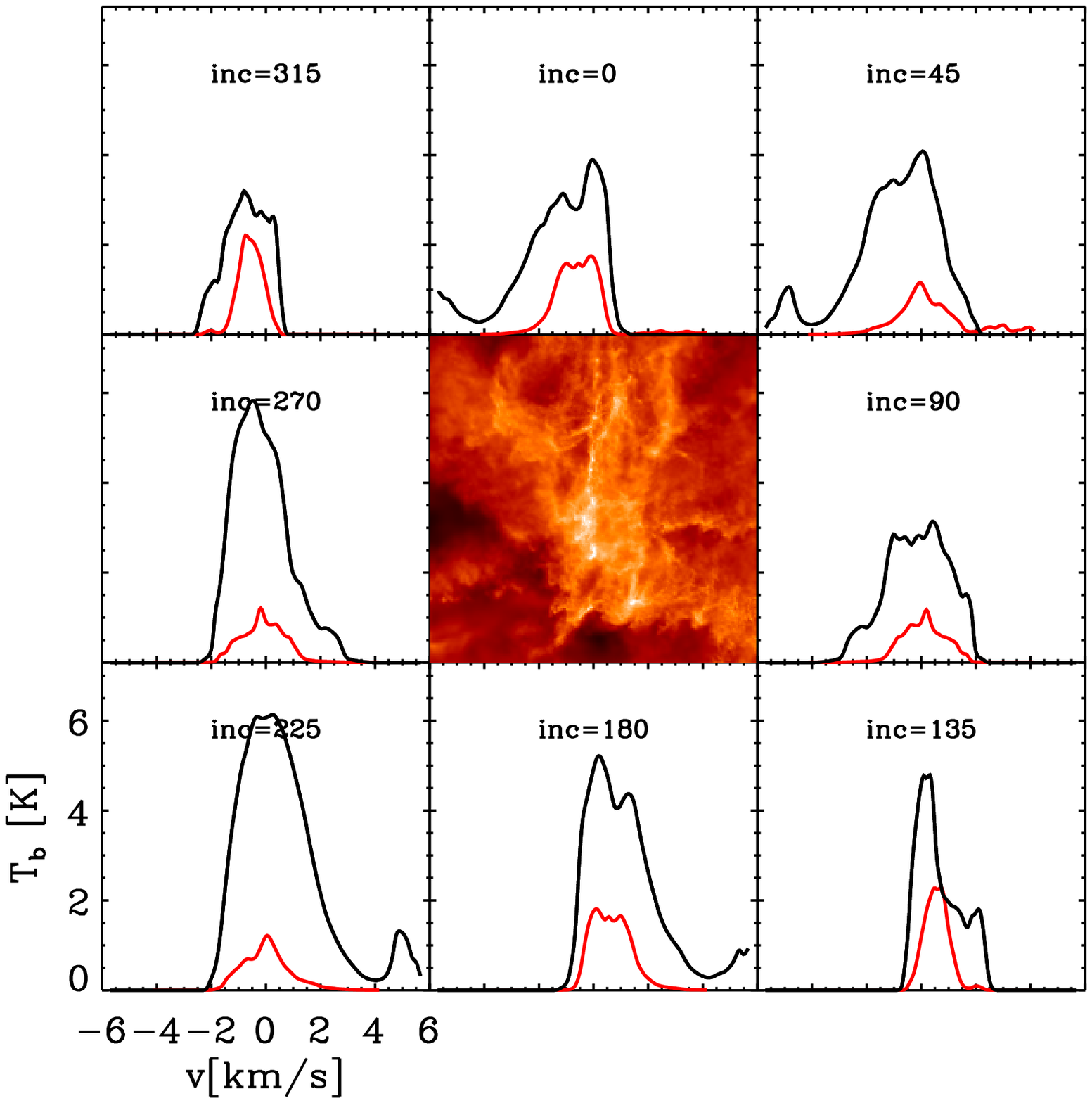}
\includegraphics[width=3.6in]{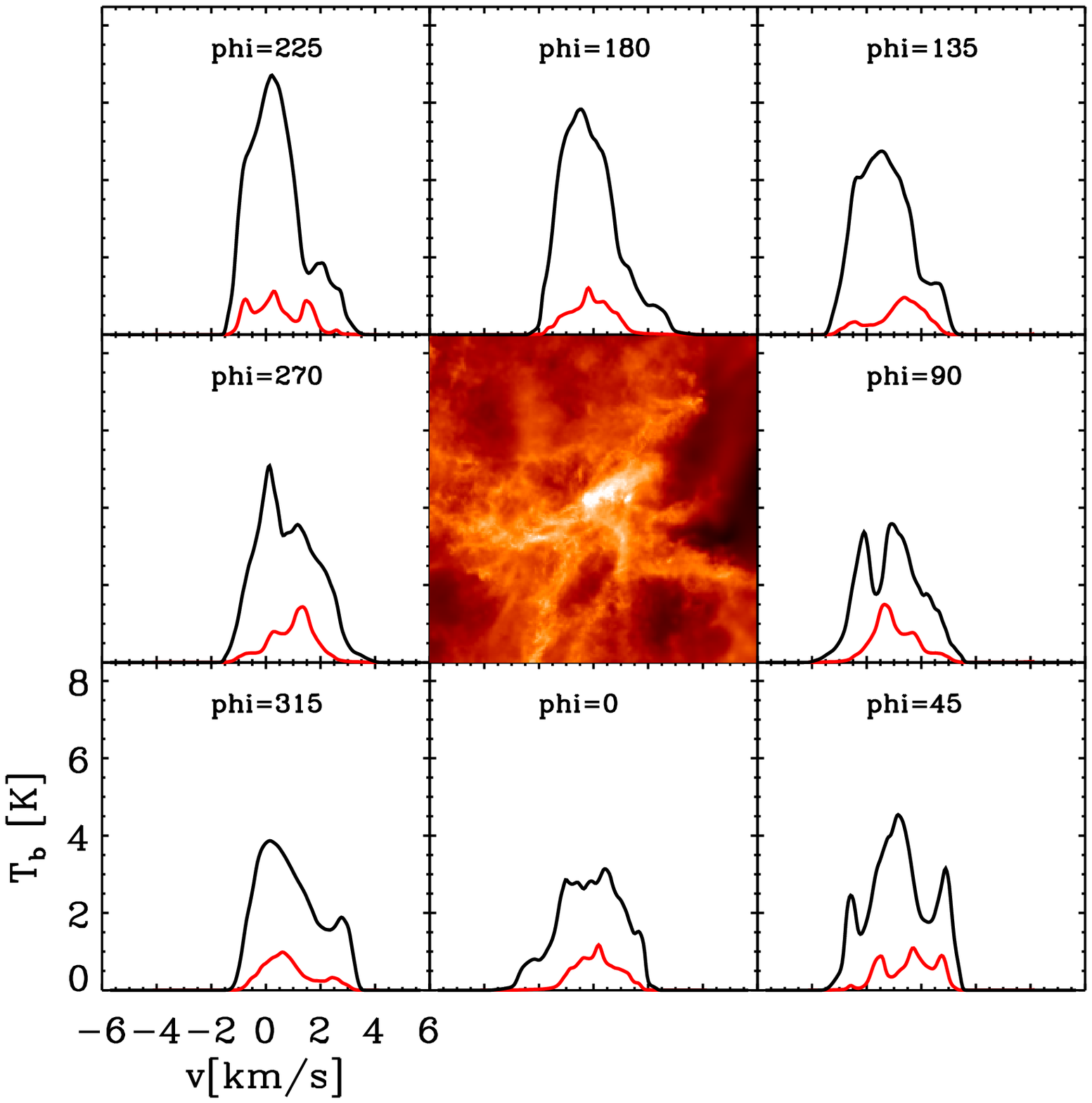}\\
\end{tabular}
\caption{The \hco (1-0) \textit{black} and N${_2}$H${^+}$ (1-0) \textit{red} line profiles from regions A \textit{top} and B \textit{bottom}. The N${_2}$H${^+}$ (1-0) is multiplied by a factor of 4 in order to be visible. The central colour image shows the column density in the plane in which the sight-lines pass through the core, and the position at which the outer panels touch the central image indicates the orientation of the sightline. The line profiles are calculated for a 0.06 pc beam centred directly on the embedded core. The central box has a physical size of 0.8 pc. In the left-hand panels the declination angle has a constant value of $\phi=0^\circ$, and in the right panels the inclination has a constant value of $inc = 90^\circ$. Note that the \n lines are symmetric through $180^\circ$ as they are optically thin.}
\label{profiles}
\end{center}
\end{figure*}

\subsection{Optically thin profiles}\label{thin}

\subsubsection{Simulation}
\fig \ref{profiles} shows the modelled line profiles along various viewing angles through the central source in Regions A and B when the sink has a mass of $ \sim 0.5$ \msun. In the left hand panels of the figures the model is fixed at, $\phi=0^\circ$, and we view the model at $45^\circ$ intervals in inclination. In the right hand panels the inclination is fixed at $inc=90^\circ$ and  we view the model at $45^\circ$ intervals in rotation. We sample a total of 14 unique lines of sight through each core centre. The red line shows the isolated \n 1-0 hyperfine line multiplied by a factor of four so that it is visible alongside the \hco (1-0) line (black). The most striking feature of the optically thin lines is that all the line profiles exhibit non-Gaussian features.

\begin{figure}
\begin{center}
\includegraphics[width=3in]{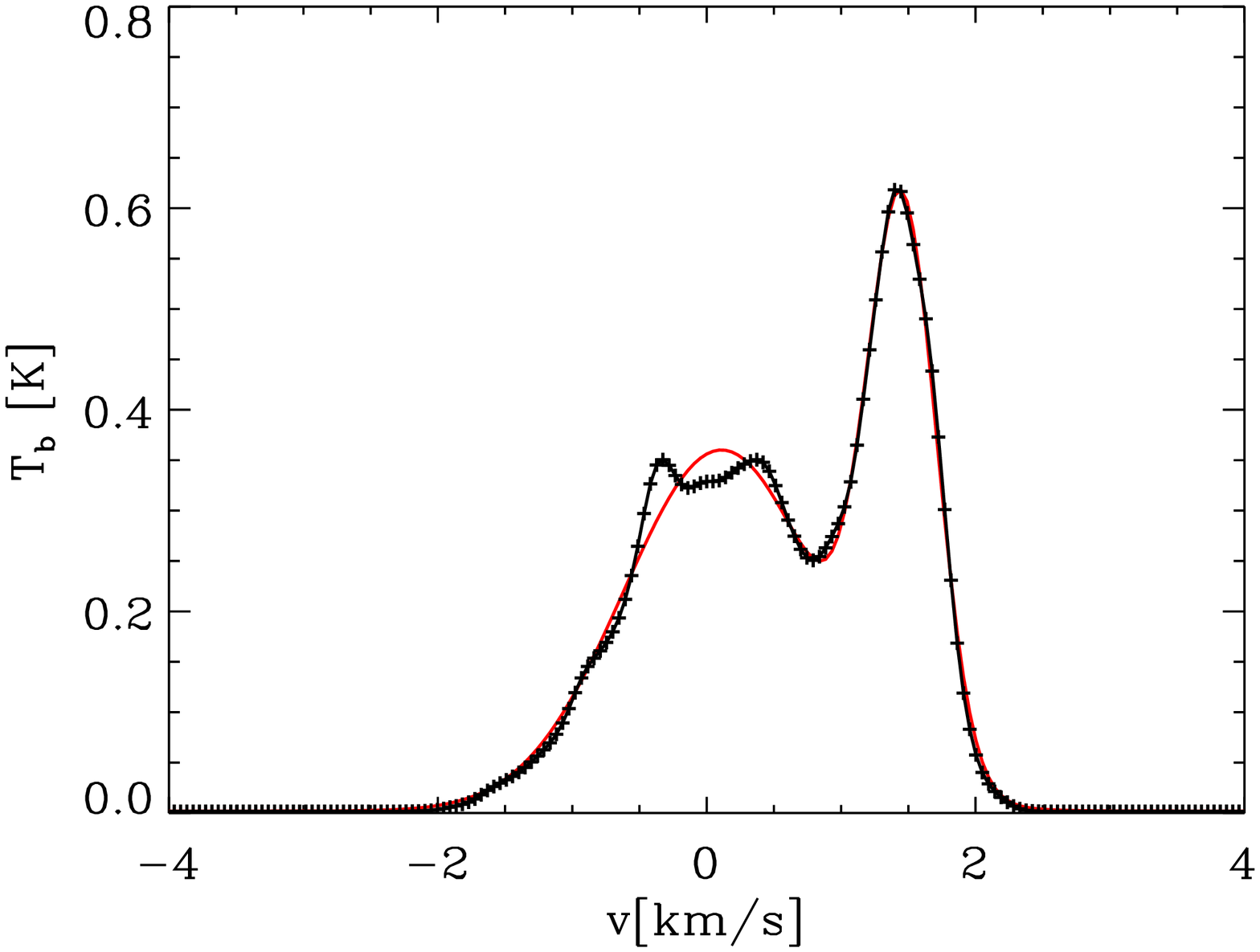}
\caption{The \n(1-0) F(2-1) line profile from Region A when viewed at $i=90^\circ$ $\phi=90^\circ$. The red line shows a simple two component Gaussian fit to the emission. The line has multiple velocity components due to a number of dense cores along the line of sight.}
\label{gausscomp}
\end{center}
\end{figure}

In Section \ref{velocities} we demonstrated that the velocity field of the region is dominated by large scale infall motions but the density field contains multiple peaks. \fig \ref{gausscomp} shows the \n line profile resulting from integrating the emission along the line of sight shown in \fig \ref{beam_density}, with a two component Gaussian fit shown in red. There are major components at velocities +1.5 \kms and 0 \kms, with the latter containing additional substructure. An examination of \fig \ref{beam_density} shows that most of the material on one side of the central protostar is collapsing towards the centre at a roughly constant velocity +1.5 \kms and that the gas traveling at this velocity contains a number of dense cores. The aggregate emission from these cores produces the linepeak at this velocity. The velocity field of the region is normalised such that the central protostar has a velocity of zero, resulting in the second Gaussian component peaking at this velocity. \fig \ref{beam_density} shows that the density field around the core is not smooth as there are dense knots of gas at either side of the core, each with a slightly different velocity. This results in the two maxima in the velocity profile in this component.

Further examination of the components of the \n lines in \fig \ref{profiles} indicates that each peak in the optically thin lines comes from a knot of dense gas in the MSFR. In our simulations MSFRs are also cluster forming regions \citep{Bonnell04,Bonnell06,Smith09b}. Such regions contain many cores of dense gas, all of which emit strongly in \n (1-0) at a velocity determined by the local speed at which gas is collapsing towards the centre. This is not true just of the simulations used here; \citet{Krumholz07}, \citet{Krumholz09} and \citet{Girichidis12a} also find multiple cores of dense gas in MSFRs. 

We note that multiple components in a line profile can also be attributed to optical depth effects, and while \n is generally optically thin, it may become optically thick in the densest regions of the core. In Figure \ref{profiles} it is immediately apparent that optical depth effects are not the causing the multiple components in the lines as the \n profiles are symmetric through $180^\circ$, an impossible outcome if the lines were optically thick. Observationally, either an estimation of the optical depth or complementary observations of another optically thin species would be required to confirm that multiple components in a line profile arise from substructure.

\begin{figure}
\begin{center}
\includegraphics[width=3in]{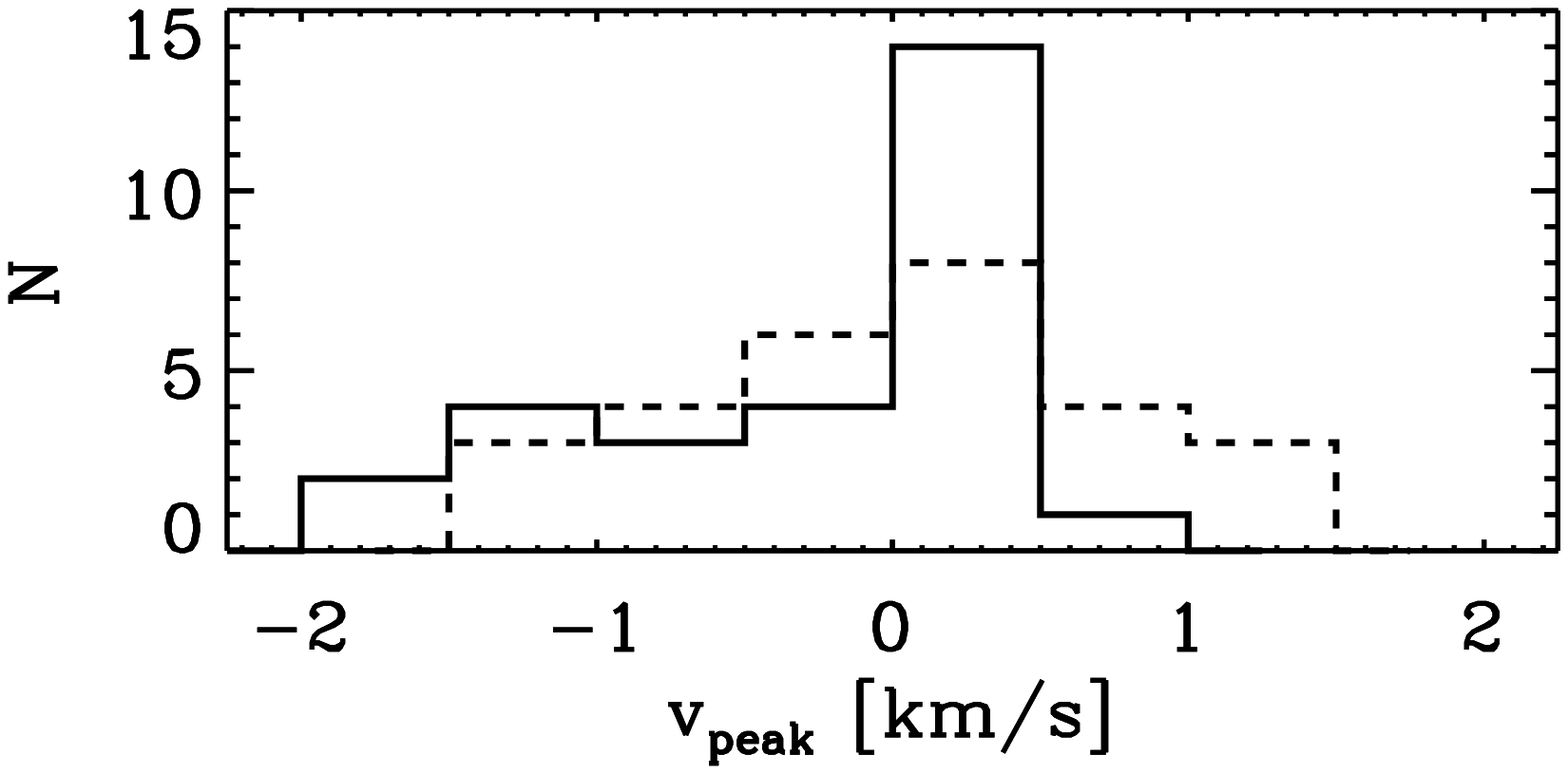}
\caption{The velocity corresponding to the peak \hco(1-0) \textit{(solid)} and \n (1-0) F(2-1) \textit{(dashed)} emission from all simulated lines of sight. The distribution of optically thick line peaks has a blue excess relative to the optically thin peaks.}
\label{velpeak}
\end{center}
\end{figure}

Another interesting feature of the optically thin emission from our MSFRs is that the line peak does not always correspond to the velocity of the massive protostar. \fig \ref{velpeak} shows a histogram corresponding to the \n line peak for all the viewing angles in \fig \ref{profiles}. The peak is frequently displaced by more than 0.5 \kms from the velocity of the massive protostar.

In all cases the observed line-widths are super-thermal, showing that bulk motions dominate the dynamics of MSFRs. Expected line-widths will be discussed more in Section \ref{time}.

\subsubsection{Observations}

While such non-Gaussianity is not observed in low mass star formation, recent studies of high mass star formation have detected multiple components from high resolution interferometer observations of cloud centres. For example, \citet{Beuther13} present an analysis of the starless prospective MSFR region IRDC18310-4 in which multiple components are clearly observed in \n (1-0) lines at the location of some of the 870 $\micron$ peaks. \citet{Csengeri11b} carried out a high resolution dynamical study of the massive clump DR21(OH) and found that while single dish \n (1-0) observations reveal a single component, the line splits into multiple components when observed with an interferrometer. This effect was particularly clear when observing the massive dense cores in the region. Multiple \n (1-0) peaks can also be seen in some cases of \citet{Fuller05} although in this case the regions are more poorly resolved compared to those studied here, an issue we will discuss in more detail later. 

In our simulations the optically thin lines have multiple components due to the presence of several cores of dense gas along the line of sight. Observational studies also find many dense cores in MSFRs \citep{Bontemps10,Longmore10,Rodon12}, suggesting that non-Gaussian optically thin lines should be a feature of massive star formation.

\subsection{Optically thick profiles}\label{thick}
\subsubsection{Simulation}

\fig \ref{profiles} shows the observed line profiles at various viewing angles through the central source in Regions A and B when it has a mass of around 0.5 \msun. The optically thick \hco (1-0) line is shown by the black line. The majority of the lines have a greater peak emission on their blue side than the red. As was shown by \citet{Zhou92} and \citet{Myers96} a double peaked profile with an excess of emission on the blue side is an indication of collapse (see Paper 1 for a detailed discussion). The profiles presented here do not always show this classical two peaked signature, but there is usually brighter emission on the blue side. The optically thick line profiles characteristically resemble a saw tooth with a sharp rise in emission on the blue side followed by a more gradual drop off and in some cases a red shoulder.

This signature is qualitatively rather insensitive to viewing angle, although in quantitive terms there are noticeable variations. In Region A collapse occurs over a very wide volume and consequently there is a peak towards the blue side in most profiles. In Region B, which forms a less massive star, the infall profiles show a greater degree of variability reflecting the fact that the inward motions are less pronounced. Still the majority of lines show a blue excess.

A more quantitative estimate of the asymmetry of the line is usually given by the normalised velocity difference $\delta V$ between the optically thick and thin components \citep{Mardones97}. However, as discussed in Section \ref{thin} since the optically thin lines cannot be consistently fit by one single component, this analysis is technically no longer valid. Nonetheless, a blue excess is seen in the optically thick lines. \fig \ref{velpeak} shows the histogram of the peak velocities in the \hco (1-0) lines. They are clearly shifted to the blue side relative to the \n (1-0) peak velocities. In 19 out of 28 cases the peak emission in \hco (1-0) occurs towards the bluewards side of the \n (1-0) peak. 

Table \ref{offsets} summarises the offsets in position between the optically thin and thick emission peaks. Table \ref{offsets} utilises two samples. Firstly, the full sample used to produce \fig \ref{velpeak}, which includes cases where the \n (1-0) emission had multiple components. Secondly, the subset of the sightlines that had only a single component in \n (1-0). In order to determine whether a sightline should be included we ignored small variations in the line profile that would be hidden by noise in a true observation, but excluded cases where there was clearly more than one major component. In this case the rest velocity of the \n emission was determined by fitting a gaussian to the profile. When compared to the typical widths of the line (several \kms) the offsets are not very large. \citet{Mardones97} required that the offset between the optically thick and thin emission be more than 0.25 times the linewidth of the optically thin component to certify a line as having a blue excess. \citet{Fuller05} found \n linewidths of order 2 \kms in their sample meaning that we should require an offset of more than $-0.5 $ \kms to classify a core as having a blue excess. Only 11 out of 28 of the \hco (1-0) emission lines have an offset of more than $-0.5 $ \kms indicating that more than half of our line profiles would not be considered infall candidates if this stricter criteria were adopted. When we exclude line profiles with multiple \n components this falls to 6 out of 18 remaining profiles.

 Our sample has a blue excess of only $E=(N_B-N_R)/N_T=(11-1)/28=0.36$ $N_B$ and $N_R$ are the number of red and blue profiles, and $N_T$ is the total number of profiles. The excess falls to 0.18 if we require the \n to have a single component. This small excess occurs despite the fact that both regions are collapsing so an excess of $E=1.0$ would be expected when considering all viewing angles.

\begin{table}
	\caption{The number of cases where the offset in \kms between the optically thick and thin (1-0) emission is of a given magnitude. We consider two samples. First the full sample where all sightlines are included. In this case the offset is between the location of the peak optically thick and thin emission. Secondly, we consider only those sightlines where the optically thin emission has no clear secondary component. In this case the offset is between the location of the peak optically thick emission and the central velocity of a gaussian fit to the optically thin emission.}
	\centering
		\begin{tabular}{ l c c c c c c}
		\hline
		\hline
		Offset & $<-1.0$  & $<-0.5$ &  $<0.0$ &$>0.0$ & $>0.5$ & $>1.0$  \\
		\hline
		 All & 3 & 11 & 19 & 9 & 1 & 1\\
	 	\hline
		Single & 1 & 6 & 15 & 3 & 1 &1\\
		\hline
		\hline
		\end{tabular}
	\tablecomments{Our sample contains 28 lines in total.}
	\label{offsets}
\end{table}

Moreover, the line profiles do not always exhibit a central dip due to self-absorption. An inspection of \fig \ref{beam_density} reveals why this is the case. This is viewed along the x-axis, which corresponds to a viewing angle inclination and declination of $i=90$, $\phi=90$ in our nomenclature. A blue asymmetric double peaked line profile relies on a collapsing region having two points at a given velocity, one at the centre of the core and the other at the outside. In Region A the outer extents of the MSFR are flowing inward with supersonic velocities of $\pm1.5$ \kms. Consequently, at velocities of $\pm 1$ \kms there is no self-absorption from gas in the outer regions and consequently all the emission from the centre of the region where the massive star forms is visible. This probably also accounts for the modest displacement between the optically thick and thin peaks. When compared to monolithic collapse with a static envelope, infall signatures should be much more difficult to detect in a chaotic cluster formation scenario.

\subsubsection{Observations}

Detecting infall towards of MSFRs has been the focus of various observational studies \citep[e.g.][]{Wu03,Sun09,Chen10}. We focus here on just two such studies: \citet{Fuller05}, which is a large scale search for infall motions, and \citet{Csengeri11} which is amongst the most highly resolved observations to date.

\citet{Fuller05} carried out a molecular line survey towards 77 candidate high mass protostars and identified 21 infall candidates, showing blue asymmetry in their \hco (1-0) lines, and 11 red profiles. The blue excess was calculated using the method of \citet{Mardones97}. The simulated observations presented here also have a large blue excess, but a greater fraction of the lines are blue and there are fewer red asymmetries. The latter is easily explained by the lack of outflows in our simulations, which could easily increase the number of red asymmetries. Another factor is that \citet{Fuller05} used a larger beam than we consider in these simulations. We will show in Section \ref{width} that this reduces the strength of the blue asymmetry. A third factor is that the blue asymmetry we observe is typically quite small in magnitude, this raises the possibility that some of the \citet{Fuller05} cores with only a slight blue excess may also be collapsing.

\citet{Csengeri11} carried out a survey of young dense cores in Cygnus X that showed the cores were dynamically evolving objects. The authors fit their observed line profiles to a model of a collapsing spherical core to estimate key properties. This model predicts that each core should have a central absorption dip. However the observed line profiles frequently did not show this feature, suggestive of large scale infall without a static envelope, as found in our simulated models.

\subsection{Time Evolution}\label{time}

\subsubsection{Simulation}

\begin{figure}
\begin{center}
\includegraphics[width=3.5in]{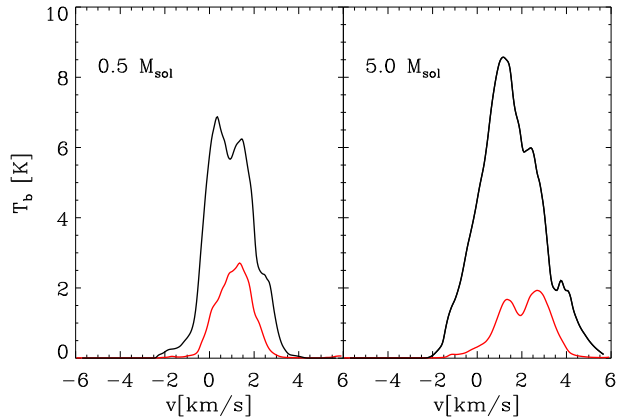}
\caption{The evolution of a line profile as the central object grows in mass in Region A. The left panel shows a central mass of $0.5$ \msun and the right a central mass of $5.0$ \msun. The black line shows the \hco (1-0) emission and the red line the \n (1-0) F(2-1) emission multiplied by a factor of four so that it is visible in the plot. Over time the line gets brighter and wider.}
\label{timeA}
\end{center}
\end{figure}

\begin{figure}
\begin{center}
\includegraphics[width=3.5in]{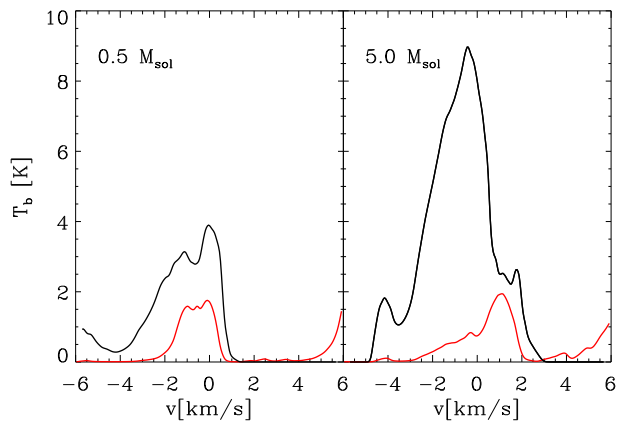}
\caption{As in Figure \ref{timeA} but for Region B. The rise in \n emission at the right edges of the plots is the neighbouring hyperfine line.}
\label{timeB}
\end{center}
\end{figure}

A further consideration is the evolution of such massive cores, as obviously not all regions will be observed when the central protostar has a mass of $0.5$ \msun as modelled above. Figures \ref{timeA} and \ref{timeB} show the evolution of the line profile observed at $i=0$, $\phi=0$ in Regions A and B when the central protostar has a mass of 0.5 \msun and 5.0 \msun. In both cases the \hco (1-0) line increases in brightness and becomes more strongly blue skewed. The \n line still exhibits non-Gaussianity.

Table \ref{timetable} shows the mean intensity peak of the lines from all viewing angles when the central protostar has mass 0.5 \msun and 5.0 \msun. The peak intensity of \hco (1-0) increases strongly in both cases, but the \n intensity is largely unchanged. Since the lines are not single Gaussians, we cannot use such a fit to determine their width, instead we find for each line the velocity range within which 95\% of the emission is contained. In both the \hco and \n case the linewidth increases as the region evolves. Table \ref{offsets_time}, however, shows that the blue excess of the sample is largely unchanged. In all cases the lines have widths in excess of the thermal scale ( $\sim 0.2$ \kms for 10K gas and $\sim 0.4$ \kms for 40K gas) as the rapid (several \kms) collapse dominates the dynamics. In all snapshots of our simulation the gas is strongly collapsing along converging filaments. Consequently we expect the effect of dynamical collapse on the line profiles to be similar throughout the entirety of the early evolution of the protostar until a HII region develops \citep[e.g.][]{Keto07,Peters12}.


\begin{table}
	\caption{Linewidths and peak line intensities}
	\centering
		\begin{tabular}{ l l l l c c }
		\hline
		\hline
		Region & Species & 0.5 \msun &  & 5.0 \msun &  \\
	 	 & & I$_{peak}$ [K] & v$_{95\%}$ [km/s] & I$_{peak}$ & v$_{95\%}$ \\
	 	\hline
		A & N$_2$H$^+$ & 0.57 & 3.31 & 0.36 & 4.19 \\
		B & N$_2$H$^+$ & 0.36 & 3.14 & 0.32 & 4.28 \\
		\hline
		A & HCO$^+$ & 6.27 & 3.95 & 8.60 & 5.05 \\
		B & HCO$^+$ & 4.63 & 4.39 & 7.59 & 5.43 \\
		\hline
		\end{tabular}
	\tablecomments{The peak line intensity, and linewidth within which 95\% of the emission is contained for the simulated line profiles. This is calculated when the central source has a mass of 0.5 \msun and 5.0 \msun respectively.}
	\label{timetable}
\end{table}

\begin{table}
	\caption{As in Table \ref{offsets} but for when the central source in each region had a mass of 5 \msun.}
	\centering
		\begin{tabular}{ l c c c c c c}
		\hline
		\hline
		Offset & $<-1.0$  & $<-0.5$ &  $<0.0$ &$>0.0$ & $>0.5$ & $>1.0$  \\
		\hline
	 	 All & 10 & 13 & 21 & 7 & 6 & 2\\
		\hline
		Single & 2 & 8 & 10 & 4 & 1 & 1\\
	 	\hline
		\hline
		\end{tabular}
	\label{offsets_time}
\end{table}

\subsubsection{Observation}

There is a general expectation that as MSFRs evolve the observed line intensity should increase due to the increased densities and temperatures, a trend that we confirm here. \citet{Wu07} showed that ultra compact HII region precursors had a lower blue excess than that that of ultra compact HII regions themselves. However, it is hard to compare these trends to the observational literature at later times than when the primary has a mass of 5.0 \msun due to contamination from outflows. For example, \citet{Chen10} carried out a survey of MSFRs identified from a survey of extended green objects \citep{Cyganowski08} and found that sources that showed stronger signs of outflows had more red line profiles.

\subsection{Higher order transitions}
\subsubsection{Simulations}

\begin{figure}
\begin{center}
\includegraphics[width=3.5in]{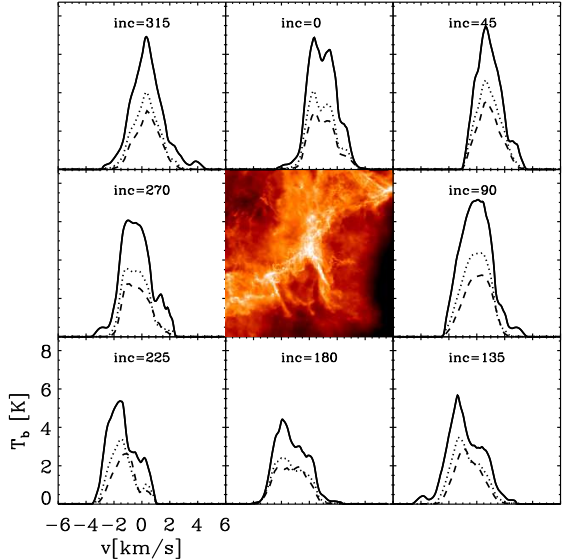}
\caption{As in Figure \ref{profiles} but showing the \hco (1-0) \textit{solid} (3-2) \textit{dotted} and (4-3) \textit{dashed} transitions for Region A.}
\label{4-3}
\end{center}
\end{figure}

Another point of interest is the line profiles of higher order transitions. In Figure \ref{4-3} we show the \hco (4-3), (3-2) and (1-0) line profiles. The lines appear more Gaussian in the higher order transitions where the critical density is higher. While the 1-0 lines frequently exhibit a red shoulder this is less pronounced as the transition number increases. Table \ref{transitions} shows the mean $\chi^2$ goodness of fit to a single Gaussian profile for each transition. As the transition increases to higher levels the $\chi^2$ statistic systematically decreases implying that a Gaussian is a better representation of the line. The lines also become less blue asymmetric with respect to the \n (1-0) peak position, particularly in the case of the (4-3) line. The (2-1) transition, which was not included in Figure \ref{4-3} has similar properties to the (1-0) transition. As expected, the peak brightness decreases due to the lower population levels at the higher transitions. The line widths also slightly decrease due to the emission originating mainly from the central regions.

The decrease in the asymmetry of the lines at higher transitions occurs because only the central dense regions have sufficient densities and temperature to excite the molecule. As we discussed in Section \ref{thick} the central dense regions have a large velocity gradient with no overlapping velocities along the line of sight. In the (1-0) case this results in a large peak with no central dip due to self-absorption. This effect is even stronger in the higher-order transitions, as there is even less chance of self-absorption from the surrounding gas. Consequently we suggest that the higher-order \hco transitions are not any more effective indicators of collapse than the lower-order transitions, and in fact in some cases might even be less sensitive to collapse motions.

\begin{table*}
	\caption{The mean $\chi^2$ goodness of fit to a gaussian profile for various transitions of the \hco line. \label{transitions}}
	\centering
		\begin{tabular}{c c c c c c c}
		\hline
		\hline
		Region & Transition & Critical Density [cm$^{-3}$] & $\chi^2$  & Blue & I$_{p}$ [K] & v$_{95\%}$ [km/s] \\
		\hline
		 A &1-0 & $1.85\E^{5}$ & $1.55 \E^{-1}$ & 8 & 6.27 & 3.95\\ 
		 A &2-1 & $1.10\E^{6}$ & $4.11 \E^{-1}$ & 8 & 4.79 & 3.79\\
		 A &3-2 & $3.51\E^{6}$ & $1.31 \E^{-2}$ & 6 & 3.83 & 3.61\\ 
		 A &4-3 & $9.07\E^{6}$ & $4.84 \E^{-3}$ & 3 & 2.94 & 3.55\\ 
		\hline
		 B &1-0 & $1.85\E^{5}$ & $1.48 \E^{-1}$ & 3 &4.63  & 4.39\\ 
		 B &2-1 & $1.10\E^{6}$ & $2.83 \E^{-2}$ & 4 & 3.25 & 3.90\\
		 B &3-2 & $3.51\E^{6}$ & $7.62 \E^{-3}$ & 2 & 2.61 & 3.31\\
		 B &4-3 & $9.07\E^{6}$ & $2.03 \E^{-3}$ & 2 & 1.92 & 3.20 \\ 
		 \hline
		\end{tabular}
		
		\tablecomments{Also shown is the number of blue profiles in each case with a blue excess of over 0.5 \kms relative to the peak velocity of the \n (1-0) line. The critical density for LTE is estimated using the relation $n_{H_2}=A_{ul}/K_{ul}$ where $A_{ul}$ is the Einstein A coefficient and $K_{ul}$ is the collisional rate coefficient at an assumed kinetic temperature of 20K. The width of the profile v$_95\%$ is the velocity range within which $95\%$ of the total emission is contained.}
		
\end{table*}

\subsubsection{Observation}

A number of studies have used the higher transitions of \hco to study massive star formation regions and successfully detected infall motions \citep[e.g.][]{Klaassen08,Roberts11,Rygl13}. The survey of \citep{Fuller05} observed the \hco (1-0) (3-2) and (4-3) transitions for all their sources. They found that the (1-0) transition had a stronger blue excess, in agreement with our findings here. Morphologically they also noted that the (3-2) and (4-3) transitions were more likely to have a single peak. \citet{Fuller05} suggested that this discrepancy might be due to the fact that there is stronger infall in the lower density outer regions of their sources. In our simulated model the velocity at which the outer regions are collapsing towards the centre is indeed higher in Region A. However, the major cause of the simpler profile is a lack of self-absorption in the dense gas. As outflows contribute the bulk of their emission to the line wings, adding outflows to our models would be unlikely to change this finding.

\subsection{Variation with beam size}\label{width}
\subsubsection{Simulation}

Our analysis so far has assumed a $0.06$ pc fwhm beam throughout, which represents the best case scenario for current observations. However, it is also useful to consider the case of larger beams, representing distant sources or single dish observation, and smaller beams, representing what might be possible with ALMA. To this end we also consider the case of a 0.4 pc ($\sim8.25\E^4$ AU) beam, equivalent to the half-width of our box, and a 0.01 pc ($\sim2\E^3$ AU) beam, which is around the typical Jeans scale in molecular clouds.

\begin{figure*}
\begin{center}
\includegraphics[width=5in]{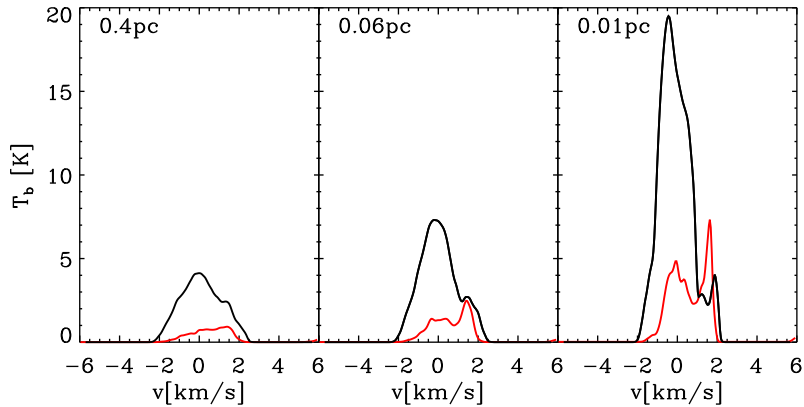}
\caption{The 1-0 transitions from Region A observed at $i=0$, $\phi=0$ with a beam fwhm of $0.4$, $0.06$ and $0.01$ pc. The black line shows the \hco line and the red the \n multiplied by a factor of four.}
\label{beam}
\end{center}
\end{figure*}

Figure \ref{beam} shows the effect of beam size on the line profile observed from Region A at $i=0$, $\phi=0$. The line brightness increases and the lines get narrower as the beam size decreases. There is also a greater discrepancy between the red and blue peaks. The \n (1-0) line becomes less Gaussian as the fwhm decreases, and very clearly contains multiple components at a fwhm of 0.01 pc. Table \ref{beamtable} shows the mean peak line intensities and linewidths for all our models as a function of beam size. In all cores and in both line transitions the  intensity increases and the line width decreases, although not to the point where the linewidth becomes thermal. 

\begin{table}
	\caption{Peak line intensity and linewidths for three beam sizes.}
	\centering
		\begin{tabular}{ l l l l c c}
		\hline
		\hline
		Region & Species & fwhm [pc] & I$_{p}$ [K] & v$_{95\%}$ [km/s] \\
		\hline
		A & N$_2$H$^+$ & 0.01 & 1.42 & 2.94 \\
		A & N$_2$H$^+$ & 0.06 & 0.57 & 3.31 \\
		A & N$_2$H$^+$ & 0.4 & 0.28 & 3.53 \\
	 	\hline
		B & N$_2$H$^+$ & 0.01 & 1.15 & 2.60  \\
		B & N$_2$H$^+$ & 0.06 & 0.36 & 3.14 \\
		B & N$_2$H$^+$ & 0.4 &  0.21 & 3.47\\
		\hline
		A & HCO$^+$ & 0.01 & 18.5 & 3.61\\
		A & HCO$^+$ & 0.06 & 6.27 & 3.95 \\
		A & HCO$^+$ & 0.4 & 4.83 & 4.48 \\
		\hline
		B & HCO$^+$ & 0.01 & 12.99 & 3.62 \\
		B & HCO$^+$ & 0.06 & 4.63 & 4.39 \\
		B & HCO$^+$ & 0.4 & 3.14 & 4.53 \\
		\hline
		\end{tabular}
	\tablecomments{The mean peak line intensity, and linewidth within which 95\% of the emission is contained for the simulated (1-0) line profiles calculated using three beam sizes.}
	\label{beamtable}
\end{table}

Table \ref{offsets_beam} shows the number of blue and red offsets observed in the regions depending on the beam size. All the regions show blue offsets between the optically thick and thin peaks with the greatest number occurring in the $0.01$ pc beam. Unfortunately, the increased prevalence of multiple components in \n complicates the interpretation of the narrow beam profiles. In Table \ref{offsets_beam} there is no entry for the single component subset because only 4 out of the 28 profiles satisfied this criteria. The \n emission is typically used to assign the rest velocity of the core and so without a unique core velocity it would be observationally ambiguous whether these were true infall profiles.

\subsubsection{Observations}
Our 0.06 pc fiducial beam size was chosen to match the resolution of \citet{Beuther13}, which uses Plateau de Bure observations. This case represents the typical resolution currently available in the literature. These studies are already revealing features hidden in older more poorly resolved surveys such as multiple components in the optically thin lines, as we have discussed above. Future observations with ALMA and the upgraded PdBI (NOEMA) should once again improve our understanding of massive star formation.

Current mm-interferometers such as the Plateau de Bure Interferometer and in particular ALMA are able to observe MSFR using a narrow beam. Therefore our $0.01$ pc beam case represents a useful test of the underlying model of star formation presented in S09. We would expect to see strong narrow peaks in \hco that are highest on the blue side of the peak \n emission component, and multiple components in the \n lines. Confirmation or refutation of these predictions should provide useful constraints on the dynamics of massive star formation. Though again we caution that outflows could change this picture.

\begin{table}
	\caption{As in Table \ref{offsets} but for different beam sizes.}
	\centering
		\begin{tabular}{ l c c c c c c c}
		\hline
		\hline
		Offset & Beam & $<-1.0$  & $<-0.5$ &  $<0.0$ &$>0.0$ & $>0.5$ & $>1.0$  \\
		\hline
		  All & 0.01 & 6 & 13 & 23 & 5 & 2 & 1\\
	 	  All & 0.06 & 3 & 11 & 19 & 9 & 1 & 1\\
		  All & 0.40 & 3 & 10 & 22 & 6 & 4 & 1 \\
	 	\hline
		 Single & 0.06 & 1 & 6 & 15 & 3 & 1 &1\\
		 Single & 0.40 & 1 & 7 & 14 & 4 &1 &1\\
		\hline
		\end{tabular}
		\tablecomments{There is no entry for the 0.01pc beam in the single component sample, as only four sightlines fulfilled this criteria.}
	\label{offsets_beam}
\end{table}

\section{Sources of Uncertainty}
\subsection{Assumption of constant abundances}\label{qual}

An important uncertainty in our method is the assumption of constant chemical abundances throughout the region. There are several reasons why this may not be true in reality. At low temperatures \hco is frozen onto dust grains, decreasing its abundance \citep{Jorgensen04}. However, this mechanism only operates below temperatures $\sim 20$ K, a temperature which is typically exceeded in the vicinity of the massive core. \n is destroyed by CO and is consequently thought to be more abundant in dense gas \citep{Bergin02}. This depends on the CO freezing out, but in hot dense gas this may not be the case. As these effects can be both positive and negative, in the absence of a full chemical model, our assumption of constant abundances is the simplest available.

We use abundance estimates from the work of \citet{Aikawa05} who applied a detailed chemical model to a collapsing Bonnor-Ebert sphere. It is unclear whether such abundances are applicable to the dense gas in massive star formation regions. Our adopted abundances are at the low end of the those found in infrared dark clouds (IRDCs) by \citet{Vasyunina11}, but are still consistent. To investigate what effect a higher abundance would have on our line profiles we run our models again with the higher abundances suggested by \citet{Sanhueza12}. For the \hco lines there is little difference, the lines are brighter but the general morphologies remain the same. However, the \n lines at an abundance of $A_{\mathrm{N2H}^+}=10^{-8.8}$ become optically thick. Since this is not what is observed in MSFRs we conclude that our original abundance of $A_{\mathrm{N2H}^+}=10^{-10}$ gives a more realistic picture. Reasons for this deviation might be that the \n should in reality be destroyed in hot regions, or that the abundance found by \citet{Sanhueza12} is only valid for the extremely cold and dense environments of IRDCs and not warmer MSFRs. If the \n is being destroyed by CO in hot regions, this might reduce the number of velocity components seen in the \n (1-0) line.

\subsection{Absence of outflows}\label{outflows}
In this paper we have concentrated on the signatures of dynamical collapse from MSFR's. However such regions may also contain outflows \citep[e.g.][]{Beuther02}. These are not included in our original simulations, and hence, their effects are not present in the modelled line profiles. Consequently our line profiles represent that which would be expected from dynamical collapse of a cluster alone. In the case of our fiducial protostellar mass of 0.5 \msun, \citet{Seifried11} showed that outflows would be confined to a column extending roughly 200AU directly above and below the protostar. Since we consider the global motions in a MSFR region 0.8pc in diameter, this should not significantly affect the resulting line profiles. 

In Section \ref{time} we consider a protostellar mass of 5.0 \msun, and in this case it expected that there should be some outflow activity. Nonetheless, even when outflows are present, the results presented here should still be observationally relevant as we are focussing on dense gas tracers. \n line profiles are almost completely unaffected by outflows as it is such a strong tracer of cold dense gas. For example, \citet{Beuther05} describes a massive core in IRDC18223-3 that shows features in the line wings of CO and CS that are indicative of outflows, however such features are entirely absent from the \n lines. Therefore, the modelled \n profiles are unlikely to be affected by outflow activity. 

However in \hco our observational comparisons may be more unreliable since most observed regions show broadened line wings from outflows. \citet{Cesaroni97} shows a protostar where the \hco line profiles have broad line wings attributed to outflows. However, the bulk of the \hco emission is clearly attributed to the dense core surrounding the protostar. This would suggest that an additional contribution may be required in the line wing, but out models should provide a good model of the behaviour of the \hco line centre. \citet{Rawlings00}, however, showed that \hco may be enhanced in the walls of a beam cavity, and that this may affect line profile morphologies \citep{Rawlings04}. even on small scales. Nonetheless, our models are the first to consider the effects of a complex gas morphology on the line profiles from massive star forming regions. If both collapse and outflows had been considered at once it would have been difficult to differentiate the two processes. Further work will be required to gain a complete understanding of \hco line profiles.

\section{Differences between low and high mass star formation}\label{comparison}
When we compare our results to those found in Paper I, several differences become apparent between high and low mass star formation. In Paper I, the optically thin \n component is Gaussian and has a narrow line width, but in the MSFR the \n line is wider and has multiple components. The mean values of $\sigma(v)$ obtained from a Gaussian fit to the \n (1-0) emission in the three low mass filaments in Paper I was $\sigma(v)=$0.28, 0.20 and 0.20 \kms. In this Paper using the fiducial beam width of 0.06 pc a similar procedure yields a mean $\sigma(v)$ of 0.80 and 0.71 \kms for Regions A and B. In Paper I the beam used was narrower (0.01 pc) as low mass star forming regions are typically closer to the observer, but unfortunately we cannot do a direct comparison for this beam size as the MSFR profiles were non-Gaussian. However, we have shown in Section \ref{width} that 95\% of the \n emission was contained over a velocity range of 2.94 and 2.6 \kms for Regions A and B. This value was only slightly lower than that seen for the 0.06pc beam.

In Paper I, the optically thick HCN (1-0) emission was highly variable with viewing angle and frequently showed no blue asymmetry (only 48\% of cores using the $\delta V$ method). In the MSFRs, the line of sight variability is slightly lower and the line profiles more consistently have blue excesses relative to the true core velocity. However, the resulting offset is not always large with respect to the width of the \n line, and the interpretation of the offsets as infall profiles is complicated by the multiple components in \n. When these factors are included the percentage of infall profiles was broadly similar in both Papers.

In the low mass cores a region of size roughly the local Jeans length (0.01-0.1 pc) is collapsing, but the gas outside this region has disordered turbulent velocities. If a line becomes optically thick in the filament, rather than in the low mass core, the core velocities will not contribute to the observed \hco line profile. In the low mass case only the dense core centre contributes to the \n emission and so there is only a single component. 

In the MSFRs the collapse motions extend over a larger region, so the lines are more likely to become optically thick in the collapsing medium surrounding the central core. However, the large velocity gradient across the region leads to little self-absorption and only a small offset between optically thick and thin components. MSFRs contain multiple dense cores, increasing the probability of detecting multiple components in optically thin lines. On the other hand, low mass star forming regions are usually not clustered, and therefore the profiles only contain a single component.

\begin{figure}
\begin{center}
\includegraphics[width=3.5in]{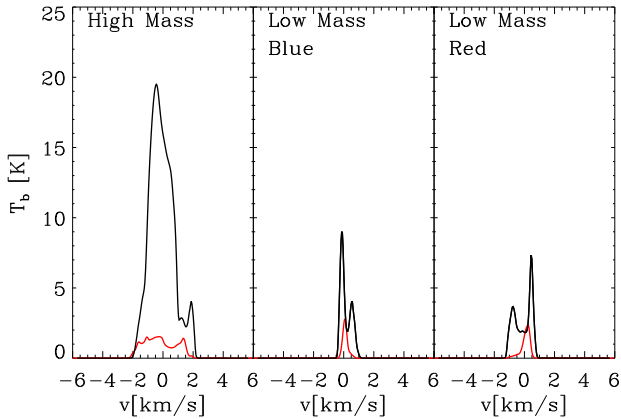}
\caption{An illustration of the \hco and \n (1-0) line profiles in high and low mass star forming regions. The \n line has been multiplied by a factor of two to make it more visible and to allow a fair comparison, we chose a beam size of 0.01 pc throughout. The high mass case corresponds to Region A viewed at i=90, $\phi=90$. For the low mass case we chose Core A from Paper I viewed at i=0, $\phi=0$ \textit{(middle)} and i=135, $\phi=0$ \textit{(right)}, which have a blue and red asymmetry respectively.}
\label{highlow}
\end{center}
\end{figure}

Figure \ref{highlow} shows a comparison between the \hco and \n (1-0) lines between a high mass and low mass core for illustrative purposes. For the high mass case, we chose Region A viewed at an inclination and declination of 90 degrees. For the low mass case we chose Core A from Paper I. To allow a fair comparison, we chose a beam size of 0.01 pc for both cases, as this is more typical of the resolution in low mass star forming regions \citep[e.g.][]{Andre07}. From many viewing angles in Paper I no clear blue asymmetry is seen, this corresponds to the case shown on the right (i=135, $\phi=0$ in Paper I) where the \hco (1-0) line is optically thick in the filament. In the middle panel we show another viewing angle in which the \hco became optically thick in the low mass core instead of the filament (i=0, $\phi=0$ in Paper I). In this case the \hco (1-0) line is brighter due to the higher core densities relative to the filament, and the line has a blue asymmetry. The \n line has a single component. In the left panel we show the high-mass case, which has the brightest \hco lines and the profile peak is to the blue-ward side of the core rest velocity. The \n line clearly shows multiple components.

\section{Conclusions}\label{conclusions}
In this paper we have carried out radiative transfer modelling of optically thin and thick line profiles arising purely from collapse motions in massive star forming regions. The underlying cloud model was obtained from the numerical simulation presented by \citet{Smith09b} in which massive stars formed at the bottom of the potential well of a proto-cluster. We assume constant abundances and only treat the early evolutionary stages of massive star formation since our simulation does not include mechanical feedback and ionising radiation. Our conclusions are the following:
   \begin{enumerate}
   
   \item \textit{Velocities:} Infall motions extend over a large volume, and there are strong velocity gradients across the modelled massive star forming regions, which are particularly steep (20 km s$^{-1}$ pc$^{-1}$ in the most massive case) across the central core. The massive star forming regions are not surrounded by a static envelope.
   
   \item \textit{Optically thin lines:} The optically thin \n (1-0) isolated hyperfine lines frequently have multiple components. This is caused by emission from dense substructure in the proto-cluster that, due to the sharp velocity gradient across the region, has a different velocity from the central core. The lines are broad, and the peak in emission does not necessarily correspond to the velocity at which the most massive star is forming.
   
   \item \textit{Optically thick lines:} The optically thick \hco (1-0) line shows only marginal blue asymmetries relative to the optically thin line peaks. The optically thick line peak is displaced by more than $-0.5$ \kms in less than half of the calculated sightlines, which is small compared to the typical observed 2 \kms linewidth observed in such regions \citep{Fuller05}. Moreover, there is rarely a central absorption dip due to the lack of a static self-absorbing envelope.
   
   \item \textit{Time evolution:} As the massive star forming region evolves its optically thick \hco lines get brighter, and both optically thick and thin line profiles get broader.
   
   \item \textit{Variation with beam:} The line brightness increases and linewidth decreases as the fwhm of the beam is reduced. In our narrowest beam the peak \hco was more frequently to the blue side of the peak \n emission. However, the \n profiles also more frequently contained multiple components in the narrowest beam. Such behaviour would be an ideal prediction to test with ALMA in order to study the dynamics of massive star formation.
   
   \item \textit{Higher order transitions:} Higher transitions of the optically thick \hco lines become increasingly Gaussian due to a growing fraction of the emission originating from a region with a large velocity gradient, where there is little self-absorption in the line. This suggests that the lower \hco transitions, namely (1-0) and (2-1) are better indicators of collapse for these regions.
   
   
   \item \textit{Comparison to low mass star formation:} The optically thick lines of MSFRs are bright, and have large linewidths. In low mass cores, the profiles are less intense and have smaller linewidths. The optically thin lines of MSFRs can have multiple peaks, due to the presence of numerous density enhancements along the line of sight.  Alternatively, optically thin lines from low mass protostars almost always have Gaussian profiles, since the ambient (filamentary) gas contributes little to the emission
   
   \end{enumerate}

\section*{Acknowledgements}
We thank Amy Stutz for her work on Paper I to which this paper is a sequel. R.J.S, R.S and R.S.K.\ gratefully acknowledge support from the DFG via the SPP 1573 {\em Physics of the ISM} (grants SM321/1-1, KL 1358/14-1 \& SCHL 1964/1-1). We are also thankful for support from the SFB 881 {\em The Milky Way System} subprojects B1, B2 and B4. 

\bibliography{../../Bibliography}
\label{lastpage}

\end{document}